# GROWTH MECHANISM AND SELECTIVE SYNTHESIS OF SWNTS

*Shuhei Inoue\*, Takeshi Nakajima, Kazuya Nomura, Yoshihiro Kikuchi*

Department of Mechanical System Engineering, Hiroshima University
1-4-1 Kagamiyama, Higashi-Hiroshima-shi, Hiroshima 739-8527, Japan
Tel/Fax: +81-82-424-5923
\* shu18@hiroshima-u.ac.jp

**ABSTRACT**

Ever since the landmark discovery of single-walled carbon nanotubes (SWNTs) in 1993, they have been considered as ideal materials for any kind of application based on their outstanding properties (e.g. mechanical strength, thermal conductivity, ultra stability, etc.), and various techniques, including laser furnace technique, arc discharge technique, and recently, Catalytic Chemical Vapor Deposition (CCVD) technique, have been developed for the high-quality macroscopic generation. Recently Hata *et al.* realized the macroscopic production after great efforts of many scientists; however, the growth mechanisms are still unclear and this incomplete knowledge prevents us from applying SWNTs to any fields. Here we can partially control the diameter distribution of SWNTs using Alcohol Catalytic CVD (ACCVD) technique and well combined catalysts. Their diameter is quite depend on the size of catalyst and their species, that is why this can be a technique to control SWNTs, and finally we propose the simple growth model.

## 1. INTRODUCTION

Single-walled carbon nanotubes (SWNTs) [1] are considered as a special material; however, their growth mechanism remains unclear. This incomplete knowledge prevents their application to any field because of the difficulty in controlling their growth. In particular, controlling the diameter and chiral indices of SWNTs are major factors in any application. In order to control their diameter and chirality, knowledge of the growth mechanism of SWNTs is essential. To date, some promising growth mechanisms have been proposed [2–5], furthermore, recently Shibuta & Maruyama [6] and Ding et al. [7] have demonstrated the nucleation of SWNTs by molecular dynamics simulation, and also Fan et al. [8] have shown the nucleation pathway for SWNTs on a metal surface by DFT calculation. In their papers carbon atoms are directly provided to the catalysts or metals. On the other hand Yudasaka et al. [9] have tried to explain the synthetic model from the experimental aspect, they have explained the nucleation of SWNTs in terms of graphitization and solubility of catalytic metals in graphite. However the growth mechanism of SWNTs has not been clear. We also proposed a simple growth model [10] that predicted the possibility of low-temperature synthesis of SWNTs using the Rh and/or Pd catalysts. Some reports [4, 11] have described the use of the Rh/Pd catalyst in a laser furnace technique; however, the synthesis SWNTs using this technique requires a reaction field in high temperature. On the other hand, the macroscopic production of SWNTs was almost realized [12]. They overcome this difficult issue of macroscopic production using the water-assisted CVD technique, so-called "Supergrowth". In this paper, we have presented the selective synthesis of SWNTs with different diameters using the combined catalysts.

## 2. EXPERIMENTAL

A schematic view of the experimental apparatus, which is similar to that of Maruyama group [12], is shown in Fig. 1. Ethanol vapor is made to flow in a quartz tube (I.D. 27 mm, O.D. 30 mm and 1200 mm) with a consumption rate of 0.9 cm3/min in liquid. The metal catalysts supported with zeolite [13] are prepared in the quartz boat, which is set at the center of an electric furnace. In the apparatus, the electric furnace consists of two parts, which can be controlled well (± 0.5%), individually. A typical procedure is as follows:

✧ Setting the quartz boat at the center of the electric furnace.
- Checking the leak-rate (usually rising from 0.4 Pa to 1.5 Pa in 5 minutes).
- Raising the temperature of the electric furnace up to the preset value, and maintaining it for 150 minutes.
- Opening the ethanol gate and flowing gaseous-alcohol for 10 minutes at 400 Pa.

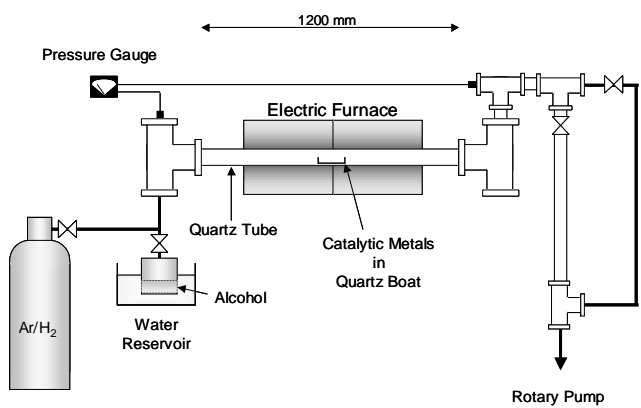

Fig. 1. A schematic view of the experimental setup.

The alcohol vessel is set in the water reservoir (10 °C) to keep the alcohol temperature constant. The temperature of the reactor, which is directly measured, rises gradually but not as high as the electric furnace temperature.

## 3. RESULTS AND DISSCUSSION

Figure 2 shows the typical Raman spectra of the raw soot, which was synthesized at 950 °C using the (a) Rh/Co, (b) Pd/Co, (c) Ni/Co, and (d) Fe/Co catalysts. The peaks in the image of each spectrum are quite similar—strong peaks around 1590 cm$^{-1}$, broad peaks around 1550 cm$^{-1}$, and a very weak peak at approximately 1350 cm$^{-1}$ called the D band. The group theory predicted that SWNTs have six Raman-active modes within their tangential G-band region [14-17], which extends from approximately 1550 cm$^{-1}$ to 1600 cm$^{-1}$. The G-band feature can be used for distinguishing the metallic and semi conducting SWNTs in a sample. Four symmetric Lorentzian features at 1553, 1569, 1592, and 1607 cm$^{-1}$ are assigned as peaks from semi conducting SWNTs, and they correspond to the $E_2(E_{2g})$, $A(A_{1g}) + E_1(E_{1g})$, $A(A_{1g}) + E_1(E_{1g})$, and $E_2(E_{2g})$ modes, respectively. On the other hand, metallic SWNTs have two strong peaks [18]: one is located at approximately 1540 cm$^{-1}$ and exhibits an asymmetric Breit-Wigner-Fano feature, which is expressed as following equation.

$$I(\omega) = I_0 \frac{[1+(\omega-\omega_0)/q\Gamma]^2}{1+[(\omega-\omega_0)/\Gamma]^2}$$

Generally, these two modes of G and D are used for simply discussing the purity of SWNTs. Figure 3 shows this G/D ratios of SWNTs obtained from the different types of catalysts, and indicates that the rhodium and palladium catalyst (Rh/Pd) exhibits poor catalysis when they are applied individually or together even in the higher temperature region (below 1000 °C it was impossible to measure this ratio). However, effective catalysts can be obtained by blending them with cobalt. In particular, the Rh/Co (▼) and Fe/Co (●) catalysts exhibit good G/D ratios. The latter is usually considered as one of the best catalysts in ACCVD. On the other hand, the Pd/Co catalyst does not show such a good G/D ratio; however, it exhibits a remarkable increase as compared to the Rh/Pd catalyst. These results indicate that in the ACCVD technique, blending cobalt can enhance the catalyst ability. Moreover, the Ni/Co catalyst (►) also shows a good G/D ratio. That is why combining cobalt with other kinds of metal might be a good catalyst for ACCVD.

Figure 4 and Fig. 5 are the low-frequency Raman spectra of SWNTs (RBM mode) where the catalyst density is different. It is speculated that the catalyst particle grows in size with an increase in its density. RBM mode is dependent on the diameter of the SWNTs, so that in this paper, we use the relationship $d$ (nm) = 248 / $\omega$ (cm$^{-1}$), where $\omega$ indicates the RBM frequency. As shown in Fig. 4, three main peaks are observed, the first peak is approximately at 216 cm$^{-1}$, second is at 254 cm$^{-1}$, and the last is at 280 cm$^{-1}$. Referring to this figure, as catalyst density becomes larger (i.e. a particle size becomes larger) a diameter distribution becomes larger.

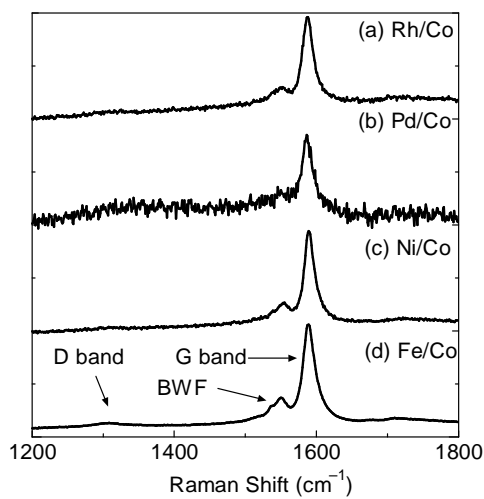

Fig. 2 Raman spectra of as-grown SWNTs grown using (a) Rh/Co, (b) Pd/Co, (c) Ni/Co, and (d) Fe/Co catalysts of 2.5 wt.% each at a furnace temperature of 950°C. The excitation wavelength is 632.8 nm.

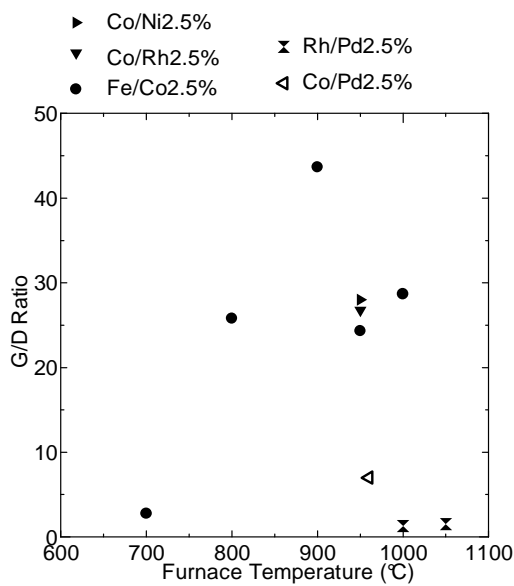

Fig. 3  G/D ratios of SWNTs obtained using different types of catalysts. G/D ratio is defined as $I_{1593}/I_{1350}$. (Excitation wavelength is 632.8 nm)
* Co/Pd catalyst could work as well as others shown in this figure, however this is quite unstable and less reproducible.

Figure 5 shows also RBM mode with different catalyst density. Basically observed phenomenon shows a same tendency with Fig. 4, and this tendency is much clear. Almost most of detected SWNTs in Fig. 5(b) are 1.3 nm in diameter, on the other hand in Fig. 5(a) there are four or five dominant SWNTs are detected from 0.9 nm to 1.3 nm. Furthermore in Fig. 5(b) two peaks are shown around 350 cm$^{-1}$ and 370 cm$^{-1}$ indicated by arrows, however since these peaks are not consistent with Kataura Plot, further experiment is essential.

In any case of these experiments shown in Fig. 4 and Fig. 5, assuming that every surface area of zeolite can support the catalyst particles, the number of particles must be equal and particle size is proportional to the density. In case of this assumption a correlation between the particle size and catalyst density is expressed as follows:

$$\left(\frac{1}{N}\rho\right)^{\frac{1}{3}} \leq r \leq \left(\frac{1}{N}\rho\right)^{\frac{1}{2}},$$

where $N$ is the number of supportable areas of zeolite, $\rho$ is catalyst density, and $r$ is particle size. According to this assumption in case of Fig. 4 a difference of catalyst density is 10 times and in case of Fig. 5 this is 2.5 times, so that the difference of diameter should be 2.15 – 3.16 times and 1.36 – 1.58 times respectively. On the other hand 1.2 nm / 0.9 nm = 1.33 times in Fig. 4 and 1.3 nm /0.9 nm = 1.44 times in Fig. 5 are experimentally shown. In case of Fig. 4 there is a big gap between ideal value and experimental value; however, in case of Fig. 5 there shows a very good agreement. It might be a reason for this disagreement in Fig. 4 that lower wave number regions ($\omega \leq 200$ cm$^{-1}$) could not be observed owing to some experimental problems. Referring to these results, changing the particle size can be an effective technique to control the diameter of SWNTs.

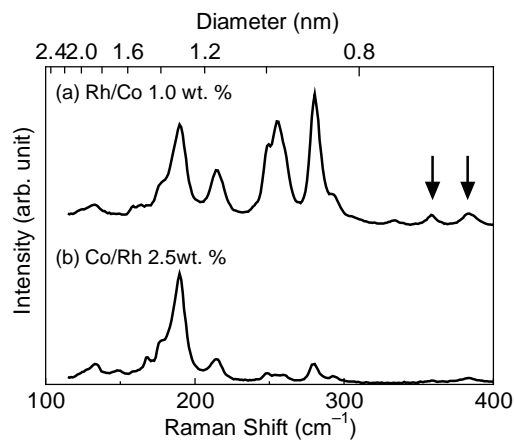

Fig. 5 The RBM mode of SWNTs with (a) Rh/Co 1.0 wt.% and (b) Rh/Co 2.5 wt.%. The excitation wavelength is 632.8 nm.

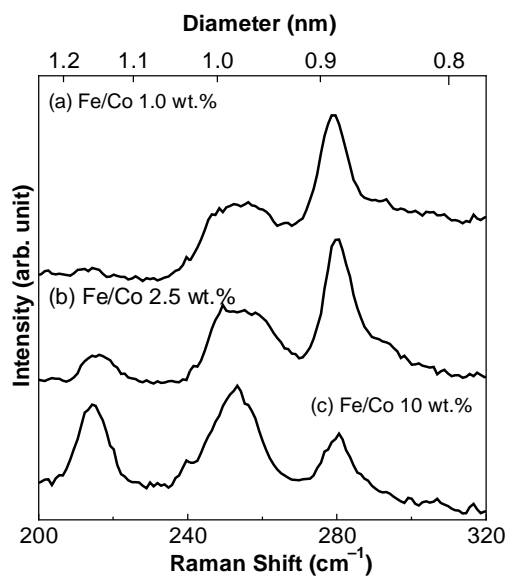

Fig. 4 The RBM spectra of SWNTs. The density of catalytic metals is different. The excitation wavelength is 632.8 nm.
  *Owing to the filter problem $\omega < 200$ cm$^{-1}$ could not be observed unlikely Fig. 5.

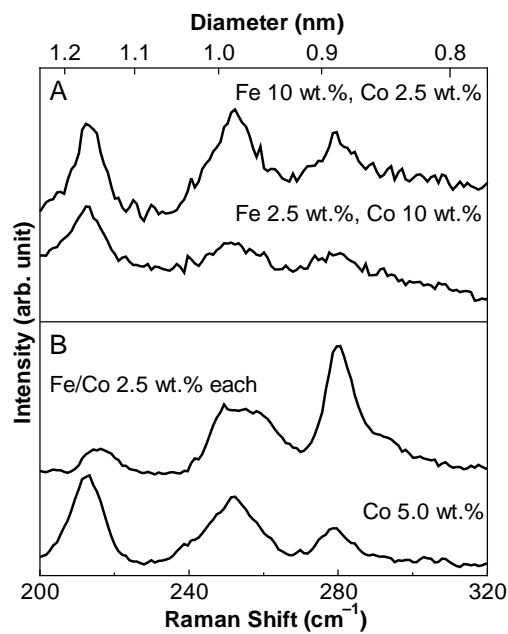

Fig. 6 The RBM spectra of SWNTs. (A) A comparison of composition between metal species. The upper is iron rich (Fe 10 wt. % and Co 2.5 wt. %), the lower is cobalt rich (Co 10 wt. % and Fe 2.5 wt. %). (B) A comparison between heterogeneous catalyst (Fe/Co 2.5 wt. % each) and homogeneous catalyst (Co 5.0 wt. %).

Figure 6(A) shows the RBM spectra of the different compositions of catalytic metal species. Since the cobalt and the iron atoms have almost same atomic masses, the size of the catalyst particles supported on zeolite may be almost equivalent. Hence, there is no influence of the size of the catalyst particle under these conditions, although the peak distribution is quite different. The RBM spectra of SWNTs synthesized on the iron-rich catalyst exhibit three peaks, which are almost equivalent. On the contrary, the RBM spectra of SWNTs from the cobalt-rich catalyst show one strong peak (216 cm$^{-1}$) and two weak peaks (254 cm$^{-1}$ and 280 cm$^{-1}$). A similar phenomenon is observed in Fig. 6(B). In this experiment, the gross weights of catalytic metals are less than those shown in Fig. 6(A), but the amount of catalytic metal is almost equivalent in both (Fe/Co 2.5 wt.% and Co 5.0 wt.%). The differences in diameter distributions in Fig. 6(B) are more distinct than those shown in Fig. 6(A). In case of the heterogeneous catalyst (Fe/Co 2.5 wt.%), the peak at 280 cm$^{-1}$ is dominant and the peak at 216 cm$^{-1}$ is very weak; however, in the case of the homogeneous catalyst (Co 5.0wt.%), the peak at 216 cm$^{-1}$ is apparently dominant. On the other hand, SWNTs are not synthesized on the iron (only) catalyst (Fe 5.0 wt.%).

These experimental results clarify the role of catalytic metals; cobalt atoms can catalyze the pyrolysis of ethanol; however, iron atoms cannot decompose ethanol. Hence, the iron (only) catalyst (Fe 5.0 wt.%) could not synthesize SWNTs. Both iron and cobalt atoms promote the growth of SWNTs; they support the core of SWNTs, as shown in Fig. 7(A), which is a schematic diagram of the growth model of SWNTs. Yudasaka *et al.* [9] mentioned the graphitization but we consider, typically, the carbon atoms tend to take the $sp^2$ structure and not the $sp^3$ structure under low pressure, especially in this synthetic technique the time scale is much longer than that of laser ablation or arc discharge that is why carbon atoms can be annealed fully. Since the electric orbits of the $sp^2$ structure spread along the x-y plane, it is difficult for carbon atoms to achieve the original cylindrical form. In order to maintain a cylindrical form against the distortion energy (Fig. 7(B)), support materials are a prerequisite. Particularly, to generate narrow SWNTs, carbon atoms must be strongly supported by more associative materials. Since the iron atoms are much associative and exhibit stronger interactions between carbon atoms than those of cobalt atoms, SWNTs synthesized on the iron-rich catalyst have narrow diameters. The relationship between synthetic temperature and diameter, which has been elucidated in some previous reports, can be easily explained through this model. At a high temperature, vibration energy is also high. Therefore, the total repulsion energy, which is the sum of the vibration and distortion energies and originated from the distortion of $sp^2$ orbits, is too large to maintain a narrow diameter. Hence, wider diameters are obtained at higher temperatures. Based on these assumptions, the simple growth mechanism of SWNTs is as follows:

First, the "core" is nucleated on the catalyst particles and supported against the distorted energy of $sp^2$ orbits. The diameter of the "core" is determined by balancing of energy in terms of size of catalyst particle, temperature, and the interaction between the catalyst and carbon atoms.

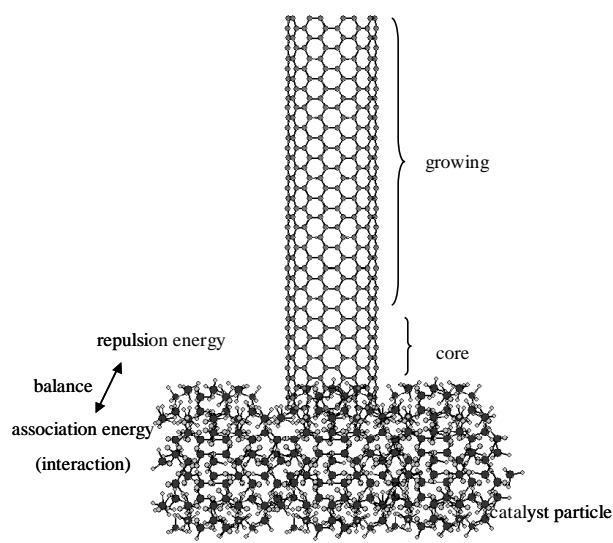

A: Growth model of SWNTs.

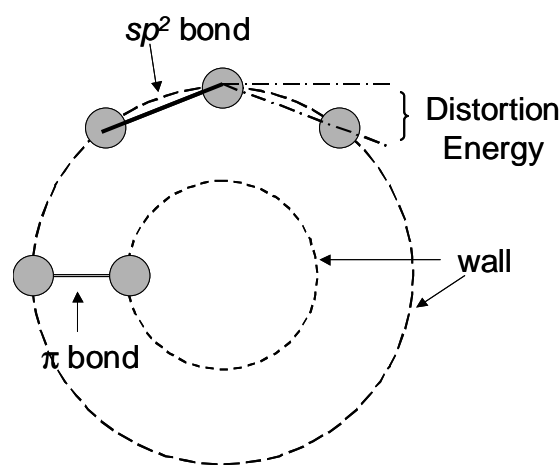

B: Energy model.

Fig. 7  A schematic diagram of the growth mechanism of SWNTs. (A) The "core" is generated and supported by the catalytic particles. (B) The repulsing energy and the stabilizing energy.

The carbon source (ethanol is used in this work) is decomposed by the cobalt atoms and supplied to the "core." The SWNTs are synthesized after some time.

This growth model can briefly explain the other synthetic technique. For example, cobalt atoms are not always required for the laser-furnace technique and arc-discharge techniques. Since carbon atoms are directly provided in these techniques, the role of catalytic metals is limited to supporting the "core." Some reports stating that using Rh/Pd catalyst may generate very narrow SWNTs are available; however, a higher temperature is required for the decomposition of ethanol and the interaction between carbon atoms. Since rhodium or palladium atoms cannot decompose ethanol at a low temperature, both or either has much stronger interactions between carbon atoms than that of the iron and cobalt atoms. Therefore, very narrow SWNTs are generated at higher temperatures. If Co/Rh or a Co/Pd catalyst is employed, narrower SWNTs can be synthesized at low temperatures. This growth model can also be extended to the growth mechanism of multi-walled carbon nanotubes (MWNTs). In the absence of a metal catalyst in the laser-furnace or the arc-discharge techniques, only MWNTs are synthesized and the synthesis of SWNTs is nearly impossible. Further, in the absence of catalyst particles, the "core" must maintain its structure by itself against the total repulsion energy, and the diameter must be larger to relax their distortion energy. Moreover, since a single layer is too weak to maintain a cylindrical structure, a pillar is needed in the center of the wide tube to enable a multi layer growth processes. MWNTs are generated following these processes with the support of their $\pi$ bonds, which are along the direction of a radius, against the distortion energy.

## 4. CONCLUSION

In this paper, the most simple growth model of SWNTs is proposed. The balance of the total repulsion energy and the association energy, which are interacted with the "core" and support catalysts, determines the diameter of the SWNTs. The explanation to the question why the dual atom catalyst is ideal for the ACCVD technique is because that one specie functions as a decomposer and the other supports the "core." According to this idea, changing the supporter can easily control the diameter of SWNTs; further, the interaction between carbon atoms is also modified.

## 5. ACKNOWLEDGEMENT

The Raman spectroscopy measurements were obtained using JRS-SYS2000 at Shimane Institute for Industrial Technology and NRI-1866M at the Natural Science Center for Basic Research and Development (N-BARD), Hiroshima University. This study was partly supported by a research grant from the Mazda Foundation.